\documentclass[sigconf]{acmart}

\setcopyright{rightsretained}
\copyrightyear{2020}
\acmDOI{}
\acmISBN{}

\acmConference[DRL4IR]{DRL4IR '20: SIGIR'20 Workshop on Deep Reinforcement Learning for Information Retrieval}{July 30, 2020}{Virtual Event, China}
\acmBooktitle{DRL4IR '20: SIGIR'20 Workshop on Deep Reinforcement Learning for Information Retrieval,
  July 30, 2020, Virtual Event, China}
\acmPrice{15.00}
\acmISBN{}
\usepackage{subfigure}
\usepackage{amsmath}
\usepackage{natbib}
\usepackage{amsfonts}
\usepackage{graphicx}
\usepackage{hhline}
\usepackage{hyperref}
\usepackage{enumerate}
\usepackage{multirow}
\usepackage{array}
\usepackage[para]{footmisc}

\usepackage[show]{chato-notes}
\newcommand{\iadh}[1]{\textcolor{black}{#1}}
\newcommand{\craig}[1]{\textcolor{black}{#1}}
\newcommand{\xiao}[1]{\textcolor{black}{#1}}
\newcommand{\xiaow}[1]{\textcolor{black}{#1}}

\newcolumntype{P}[1]{>{\centering\arraybackslash}p{#1}}

\usepackage{marginnote}
\begin{document}
\fancyhead{}
\title{Deep Reinforced Query Reformulation \\for Information Retrieval} 



\author{Xiao Wang}
\affiliation{\institution{University of Glasgow}}
\email{x.wang.8@research.gla.ac.uk}

\author{Craig Macdonald}
\affiliation{\institution{University of Glasgow}}
\email{craig.macdonald@glasgow.ac.uk}

\author{Iadh Ounis}
\affiliation{\institution{University of Glasgow}}
\email{iadh.ounis@glasgow.ac.uk}



\settopmatter{printacmref=false}

\begin{abstract}


Query reformulations have long been \iadh{a key mechanism} to alleviate the vocabulary-mismatch problem in \iadh{information retrieval}, for example by expanding \iadh{the queries with} related query terms or by generating paraphrases of the queries.  In this work, we propose a deep reinforced query reformulation \xiao{(DRQR)} \iadh{model} \xiao{to automatically generate \iadh{new} reformulations of the query. To encourage the model to generate queries which can achieve high performance when performing the retrieval task, we incorporate query performance prediction into our reward function. In addition, to evaluate the quality of the reformulated query in the context of information retrieval, we first train our DRQR model, then apply the retrieval ranking model on the obtained reformulated query. Experiments are conducted on the TREC 2020 Deep Learning track MSMARCO document ranking dataset. Our results show that our proposed model outperforms several query reformulation model baselines when performing retrieval task.} \xiao{In addition, improvements are also observed when combining with various retrieval models, such as query expansion and BERT.}



%
%
 

\end{abstract}


\maketitle

\section{Introduction}
Vocabulary mismatch is an inherent problem \craig{in} information retrieval \iadh{(IR)} \craig{tasks}, due to the \iadh{possible} inconsistency between the way users express their information needs and the \craig{manner in which relevant content is} described \iadh{in the documents}.
In order to alleviate this vocabulary mismatch problem \iadh{in IR}, many approaches \craig{have been} proposed. For instance, \craig{in relevance feedback, additional terms identified from known relevant documents are added to \iadh{the} original user's query; pseudo-relevance feedback (PRF) is the name given to the automatic process, where the original query is reformulated (typically expanded) using terms occurring in the pseudo-relevant set of documents \xiaow{--} typically the top-ranked documents in response to the initial query~\cite{croft2010search}.}

\craig{More recently, there has been a move towards addressing more complex information needs where \iadh{the} user queries are \iadh{often expressed} as questions rather than ``keywords''. Indeed, recent context-aware neural ranking techniques such as BERT have \iadh{been shown to be effective over} question-like queries~\cite{dai2019deeper}. The research by participants in the recent TREC 2019 Deep Learning track~\cite{craswell2020overview} exemplifies recent work in this direction.}
\craig{To address the vocabulary mismatch for question-like queries, we are inspired by the work of Zerveas et al.~\cite{zerveasbrown}, who aimed to learn how to generate {\em paraphrases} (alternative question formulations) of queries using a deep learned text generation model called} {\em query2query}.


\iadh{Indeed, in} recent years, deep neural networks \craig{have played} an important role in text processing-related tasks. For instance, \craig{sequence to sequence (seq2seq) models (based on recurrent neural networks, RNNs) have \iadh{demonstrated an} ability} to learn the meaning of a sentence. Seq2seq models have been extensively used, for instance, to generate \craig{paraphrases} of an input sentence~\cite{yang2019end}; to simplify natural language \craig{queries} into a keyword query~\cite{liu2018generating} or to extract the key phrases of \craig{a} given input document~\cite{yuan2018one,chen2018keyphrase}. 


However, the traditional sequence to sequence (seq2seq) model suffers from two problems: the exposure bias and the inconsistency between the train and test measurement metrics~\cite{ranzato2015sequence,keneshloo2018deep}. 
\craig{To address these problems, reinforcement learning has been applied to sequence to sequence modelling, such that the \iadh{RNN-based} seq2seq model is regarded as an {\em agent}, while} an {\em action} is generated by a stochastic policy based on the reward given by the reward function~\cite{ranzato2015sequence,keneshloo2018deep}.
\xiao{In this work, we \craig{propose} the Deep Reinforced Query Reformulation (DRQR) model, which is a deep reinforcement learning-based seq2seq model \craig{that} can automatically generate query reformulations for a given input query.} 
\craig{
The reward function in our reinforcement learning setup is inspired by previous work in {\em selective} pseudo-relevance feedback~\cite{he2007combining}: indeed, the effectiveness of pseudo-relevance feedback is sensitive to the quality of the initial retrieval results~\cite{carpineto2012survey}, 
and therefore {\em query performance predictors}~\cite{he2006query,carmel2010estimating} can be used to identify when it suitable to apply PRF~\cite{he2007combining}. 
Similarly, we use query performance predictors within the reinforcement reward function to select high quality paraphrases -- in doing so, the predictors help the learning algorithm to produce paraphrases that are predicted to be effective, and helps to bridge the gap between sequence prediction (the training task) and retrieval effectiveness (the ultimate ``test'' task).}
\craig{In summary, this paper provides three contributions:} (1) We employ \iadh{a} reinforcement learning technique \craig{within} our \xiao{query reformulation model} \craig{to generate query reformulations}; (2) the model incorporates the query performance prediction into our reward function to direct the learning towards good query reformulations; (3) We demonstrate the effectiveness of our reinforcement learning-generated query paraphrases within a state-of-the-art BERT ranking model upon the TREC 2019 Deep Learning track test collection.

\looseness -1 The remainder of this paper is structured as follows: In Section~\ref{sec:related}, we position our model 
with respect to the related work.  Section~\ref{sec:model} presents our proposed deep reinforcement learning model. Research questions and experimental setup are \iadh{described} in Sections~\ref{sec:RQ} \& \ref{sec:exp}. Results \iadh{analysis} and conclusions respectively follow in Sections~\ref{sec:results} \& \ref{sec:con}.
\section{Related Works}\label{sec:related}
We consider two aspects of related work, namely, \iadh{a review of relevant information retrieval (IR) approaches addressing the vocabulary mismatch problem}, query performance predictors, and \iadh{work related to} text generation models.

\subsection{Paraphrasing Queries}

Many approaches have been proposed to alleviate the vocabulary mismatch problem by adjusting the formulation of the query, including automatic pseudo-relevance feedback \iadh{techniques}, \iadh{ranging} from Rocchio\iadh{'s algorithm}~\cite{rocchio1971relevance} to \iadh{the DFR relevance feedback approaches}~\cite{amati2002probabilistic} \iadh{through} relevance models such as RM3~\cite{lavrenko2001relevance}. Such query expansion approaches typically reweight the query terms, such that new \iadh{query} terms may be added with non-binary weights.

\craig{Alternatively, generating paraphrases of user queries has been proposed to address the} ``lexical chasm'' problem. Many \iadh{studies} \craig{have employed} lexical paraphrases of queries to expand the original query thus improving the retrieval performance. For instance, Zukerman et al. \craig{used an} iterative process to identify \iadh{similar} phrases \iadh{to} a query \iadh{based on} WordNet~\cite{zukerman2002lexical}. \craig{However, static resources such as WordNet \craig{may not be able to address the changing nature of search}\xiao{, for example the new words.} 
\craig{One \iadh{recent} branch of work involves considering previous user queries for sources of reformulations. For instance,} Jones et al. generated candidate substitutions for phrases of the input query based on logs of previous user queries}~\cite{jones2006generating}. 
Later, \emph{Statistical Machine Translation} (SMT) \craig{techniques were} employed to expand the query by first generating the phrase-level paraphrases of the query, then selecting terms from the n-best paraphrases queries as expanded terms~\cite{riezler-etal-2007-statistical}. For instance, \iadh{a} query \iadh{such as} ``paint a house'' is rephrased as ``decorate a room'', \iadh{where} the terms ``decorate'' and ``room'' can be used to expand the original query. However, these methods are not neural \iadh{models-based} and \iadh{require extensive efforts on users} to select from the rephrased phrases. 


\looseness -1 Another promising \iadh{approach} is to expand the original query by generating the query-level paraphrases at once while preserving the meaning of the original query. For example, ``do goldfish grow'' and ``how long does a goldfish grow'' form a pair of paraphrases. \iadh{Zerveas et} al.~\cite{zerveasbrown} \craig{proposed} a \emph{query2query} method based on the Transformer model to generate three rephrased queries given the input query. Then the three generated paraphrases together with the original query can be used to retrieve relevant documents, \iadh{with the aim of enhancing the retrieval effectiveness}. However, \iadh{Zerveas et al.} did not intervene in the process of generating the paraphrases, \iadh{meaning that} their paraphrase generation model failed to consider \iadh{the} generated paraphrases' \iadh{retrieval effectiveness}. 
In \iadh{this work}, our model takes the query \iadh{retrieval} performance into consideration while generating the paraphrases of a given query. 
\subsection{Query Performance Prediction}


\craig{A risk when generating paraphrases of queries is that they might not be \iadh{of} high quality, and lead to degraded retrieval effectiveness. To address this, in this paper, we make use of query performance predictors (QPP).} The goal of query performance prediction is to predict the search effectiveness \craig{for a given query} in advance, \craig{without any} relevance information \craig{provided by humans}. \craig{Query performance prediction has been used to apply different retrieval approaches for queries that are predicted to be difficult} \xiaow{--} 
for instance, selective query expansion exploits \iadh{query} performance \iadh{predictors} to decide whether to expand the original query or not~\cite{cronen2004framework,he2007combining}. \xiao{Furthermore, Lv et al.~\cite{lv2009adaptive} suggested to use query performance prediction to decide the number of additional terms to expand a given query with when performing pseudo relevance feedback. However, both of these approaches using query performance predictors are more focused on expanding the original query with additional terms rather than generating an entire paraphrase of the query at once, as we apply in our work.}

Query performance prediction approaches \craig{can mainly be categorised} as being {\em pre-retrieval} and {\em post-retrieval} \craig{in nature}, where \craig{pre-retrieval predictors only} exploit the raw query and statistics of the query terms, as recorded at indexing time. In \craig{contrast}, post-retrieval \craig{predictors} analyse the retrieved documents, in terms of score distributions and/or content. Based on this, our work uses pre-retrieval predictors as a reward signal to \craig{generate query paraphrases that are expected to be effective}. 

\subsection{Text Generation Models}


\xiao{Neural text generation models have achieved outstanding \iadh{performances} in many applications. In this paper, we cast our query reformulation task as a form of text generation task}\xiaow{,} which can be 
\craig{
addressed using sequence-to-sequence models (seq2seq). Below, we review seq2seq \iadh{models} and discuss how they can be enhanced using reinforcement learning.}

\subsubsection{seq2seq \iadh{models}}
Sequence to sequence models~\cite{sutskever2014sequence} generally \iadh{consist} of \craig{an} RNN-based encoder and decoder. The encoder encodes the input sequence into a \craig{fixed-size} hidden vector, based on which the decoder generates the predicted sequence. However, an information bottleneck \craig{can form when trying} to encode all the information of the source sequence into a single vector. Later, an attention mechanism \craig{was} proposed by Bahdanau et al.~\cite{bahdanau2014neural} and Loung et al.~\cite{luong2015effective} to allow the decoder to build \craig{a} direct connection with the encoder and to focus on a particular part of the source \craig{sequence at} each decoding step. 
\craig{Later}, Gu et al.~\cite{gu2016incorporating} proposed the copy mechanism, which is a mixture of generating a token from the vocabulary or copying a token from the source sequence. The copying mechanism enables a seq2seq model to generate out-of-vocabulary words in the target, by selecting \xiaow{words} from the \craig{source} sequence. 

The sequence to sequence models have been used for a variety of IR tasks. For instance, Sordoni et al. applied a hierarchical \iadh{RNN-based} model to generate query suggestions~\cite{sordoni2015hierarchical}. Liu et al. transformed the natural language query into the keyword query~\cite{liu2018generating} \iadh{with the aim} to improve the \iadh{retrieval} effectiveness \iadh{of} term-matching IR models. In addition, in the work of He et al.~\cite{he2016learning}, a seq2seq model is trained to reformulate the input queries,
and the beam search technique is employed to generate multiple query reformulations as candidates, from which good reformulations are selected by a candidate ranking process. Considering that time-complexity \craig{is increased} by beam search, in our work, we \craig{build our query reformulation model based on a seq2seq model that includes both attention and the copy mechanism}. To encourage our reformulation model to reformulate the original query using different words. we adopt the \emph{one2many} technique in Catseq~\cite{yuan2018one}. \iadh{The Catseq} model \iadh{has been} originally designed to deal with the keyphrase generation 
problem by generating multiple keyphrases \craig{conditioned} on the input text. \xiao{Instead of using the \iadh{above} generation technique, for example \iadh{the} beam search technique,} the \iadh{Catseq} model concatenates multiple generated phrases \iadh{into} a sequence as output to achieve the diversity goal.
\xiao{We build our \iadh{proposed} model based on Catseq, where each word of the ground-truth paraphrase is regarded as a one-word keyphrase and the input query is regarded as the input text.} 






\subsubsection{Reinforcement learning for text generation}\label{ssec:related:RL}
\craig{While} traditional sequence to sequence models are trained using \xiao{the word-level} cross-entropy loss \xiao{function},
\craig{their usefulness may only be determined for some information retrieval \iadh{tasks}, which would be evaluated using different metrics. Indeed, in our query reformulation task, we may consider reformulation success in terms of retrieval effectiveness, but typical retrieval metrics are not differentiable \xiao{with respect to the model parameters,} 
and hence cannot be considered within the seq2seq learning process.} \craig{Further, traditional sequence to sequence models suffer from {\em exposure bias}, in that during training they are fed the ground truth tokens one at a time \xiaow{--} this creates models that are conditioned based on \iadh{the} correct words~\cite{ranzato2015sequence}, and as a results produce less accurate generations at test time.}

To avoid these problems, reinforcement learning has been applied to a wide array of text generation tasks, \craig{including}, keyphrase generation~\cite{chan2019neural}, summarisation~\cite{paulus2017deep} and paraphrasing~\cite{li2017paraphrase}. Buck et al. proposed a question reformulation model based on seq2seq trained using reinforcement learning for the QA task~\cite{buck2017ask}. The reward is calculated based on the returned answer in response to the reformulated question. Different \craig{from} their work, our target is document retrieval rather than question-answering. In addition, Nogueira et al.~\cite{nogueira2017task} proposed a reinforcement \iadh{learning-based} query reformulation model that selects expansion terms from the top-ranked documents returned by the initial retrieval. Their reward function is designed to \iadh{leverage} recall when conducting retrieval on the predicted query sequence at the end of each episode. However, similar to pseudo-relevance feedback, the model is sensitive \iadh{to} the initial retrieval \iadh{performance}. In addition, due to the use of recall in their reward function, \craig{the need to retrieve at each iteration means it takes a considerable} time to train the RL model -- indeed, they report training for 8-10 days\footnote{\craig{Despite significant efforts, we were unable to get the code provided by Nogueira et al.~\cite{nogueira2017task} to run on modern GPU hardware, a problem acknowledged by the authors.}}. 



In our work, we cast our \craig{query reformulation} learning task as a reinforcement learning problem and employ the policy gradient algorithm REINFORCE~\cite{williams1992simple}. \craig{Concretely}, we adopt the Self-Critic (SC)~\cite{rennie2017self} REINFORCE model to reduce its high-variance. The goal of our \iadh{proposed} model is to improve the effectiveness of the \iadh{retrieval} task. However, different from existing work, our RL approach incorporates the rewards not only from the lexical match between the generated sequence and the source sequence but also from a \craig{retrieval-related reward}, \craig{obtained from the query performance predictors}. Indeed, by using query performance predictors to guide the paraphrase generation instead of retrieval recall (as used by Nogueira et al.~\cite{nogueira2017task}), \iadh{this} results in faster training time, as the predictors only require collection statistics. 

%





\section{A Deep Reinforced Query Reformulation Model}\label{sec:model}
 
\begin{figure}[tb]
  \centering
  \includegraphics[width=\linewidth]{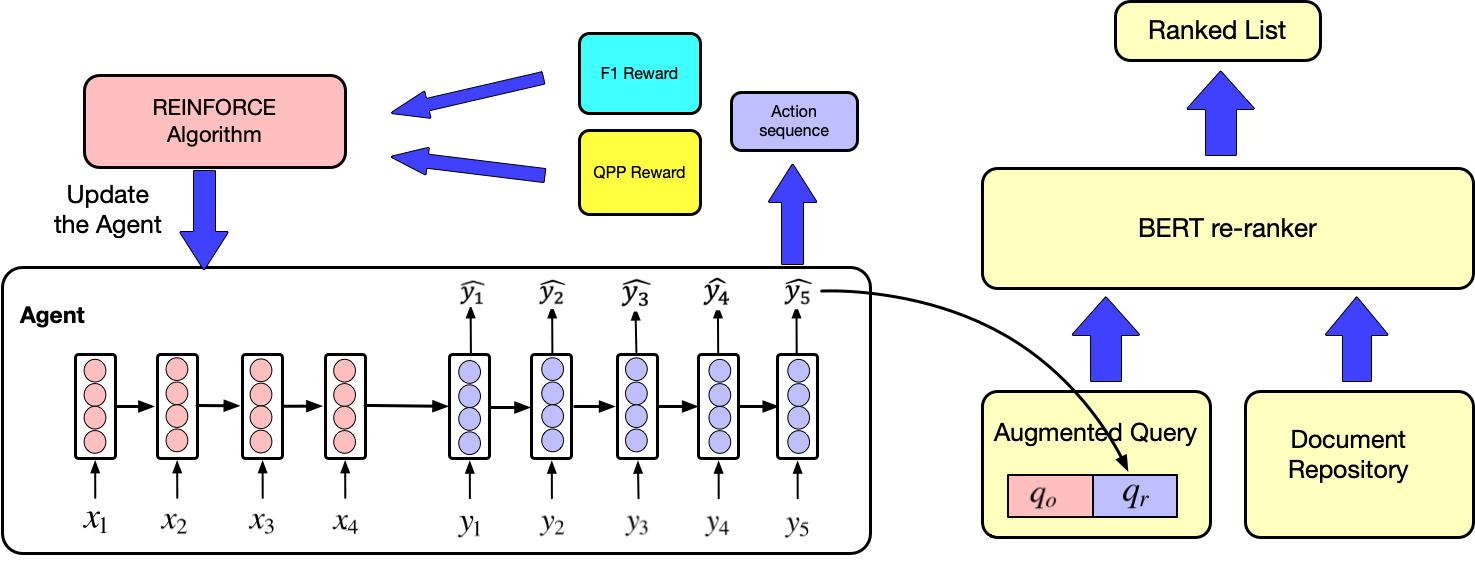}\vspace{-0.5\baselineskip}
  \caption{Architecture of our proposed Deep Reinforced Query Reformulation (DRQR) model. }\vspace{-\baselineskip}
  \label{fig:framwork}
\end{figure}
In this section, we describe our  Deep Reinforced Query Reformulation (DRQR) model in detail. We first formally define our problem in Section~\ref{ssec:probem defin} and the detailed training process is explained in Section~\ref{subsec:traning process}. Our reward function is specified in Section~\ref{subsec:reward function}.

\subsection{Query Reformulation Problem Definition}\label{ssec:probem defin}
\looseness -1 Formally, the task \craig{performed by the} DRQR model can be \craig{described} as
follows: \craig{given a pair of input user query $X=[x_1,x_2, ..., x_N]$ \craig{of} \iadh{ $N$ terms length} and a paraphrase of that query $Y=[y_1,y_2, ..., y_M]$ with length $M$, the model is trained to produce a reformulation $\hat{Y}=[\hat{y_1},\hat{y_2},...,\hat{y_M}]$. This predicted query  $\hat{Y}$ should aid a retrieval system to identify documents that are relevant to the original query $X$. } 


\subsection{Training Process}\label{subsec:traning process}
In this section, we describe the training process of our DRQR model. 
Figure~\ref{fig:framwork} presents the \craig{architecture} of our model, which consists of two parts: The left part is the query reformulation model, which is trained using the REINFORCE algorithm. After the query reformulation model is trained, the obtained reformulated query together with the original query form \iadh{an} augmented query. The right part is the retrieval pipeline, \iadh{which} scores the documents based on the augmented query. 
We first introduce the backbone text generation models: a seq2seq model with \iadh{an} attention and copy mechanisms, then the reinforced learning process.
\subsubsection{Encoder-decoder model}
Our query reformulation model adopts the recurrent neural network \iadh{(RNN)-based} encoder-decoder framework. Generally speaking, the encoder encodes the input sequence into a fixed-length vector representation, then the decoder decodes \iadh{it} back into a variable-length sequence. We adopt bi-directional  gated recurrent units (GRU-RNN) as the encoder~\cite{cho2014learning}, which reads each word, then updates the hidden state: 
\begin{equation}
  h_{n} = GRU_{encoder}(h_{n-1},x_n)
\end{equation}
Thus the encoder converts the input sequence into a sequence of real-valued \xiao{vectors}:
$H_e=[h_1, h_2, ..., h_N]$.

The decoder is an uni-directional GRU model, which is trained to generate the current hidden state conditioned on the current word $y_m$ and the previous hidden state: 
\begin{equation}
    s_{m} = GRU_{decoder}(s_{m-1},y_m)
\end{equation}

\craig{An} Attention mechanism~\cite{bahdanau2014neural} is used to \craig{determine} the importance \craig{of} each word from \iadh{the} source sequence given the current decoder hidden state $s_m$ when generating token $y_m$. \craig{At} each decoder step $t\in [1,M]$, we have the encoder hidden states $H_e=[h_1, h_2, ..., h_N]$ 
and \iadh{the} current decoder hidden state $s_t$, then we get the attention scores by applying a single-layer feed forward attention function: 
\begin{equation}
    {e}^{t}=\left[{s}_{t}^{T} {h}_{1}, \ldots, {s}_{t}^{T} {h}_{N}\right]
\end{equation}

To address the importance of each word from the input sequence, the softmax function is applied to the obtained attention scores. Then we get the attention distribution  $\alpha^{t}$, which is the probability distribution of the input sequence\craig{, as follows:}
\begin{equation}
    \alpha^{t}=\operatorname{softmax}\left(e^{t}\right)
\end{equation}

Finally the attention weights \craig{are} used to represent the encoder hidden states as a context vector: 
\begin{equation}
    {c}_{t}=\sum_{i=1}^{N} \alpha_{i}^{t} {h}_{i}
\end{equation}

\craig{Next, an effective query reformulation often involves using at least one of the input query words appearing in the reformulated query. This contrasts with other conventional seq2seq tasks, such as machine translation, where it is rarer for the same words to appear in both input and output. To address such a need, Gu et al.~\cite{gu2016incorporating} proposed a copy mechanism, which we also adopt in this work.} At each generation step $t$, \craig{the copy mechanism decides} to switch \craig{between} generating words from the vocabulary or copying words from the input source sequence. 
\begin{equation}
    p\left(\hat{y_t}\right)=q_t \cdot p_{p}\left(\hat{y_t}\right)+\left(1-q_t\right) \cdot p_{g}\left(\hat{y_t}\right) \label{eqn:copy}
\end{equation}
where $q_t$ is conditioned on the context vector and the decoder hidden state and decides \craig{to} switch between \iadh{the} generation or copying \craig{modes}.
We employed the teacher forcing algorithm~\cite{williams1989learning} to train the model \craig{using the ground-truth $Y=[y_1,y_2, ..., y_M]$}. The  maximum-likelihood training objective can be described as:
\begin{equation}
\mathcal{L}(\theta)_{ML}=-\sum_{t=1}^{M} \log p(y_{t} | y_{1}, \ldots, y_{t-1};\theta)
\end{equation}
where $\theta$ \craig{denotes} the \craig{parameters} of the \xiao{seq2seq models}.
\craig{However, as mentioned in Section~\ref{ssec:related:RL},} minimising the maximum-likelihood loss function \craig{may not necessarily lead to generated query reformulations that are effective in nature.} Thus, there is a discrepancy between the training objective and the overall objective. In addition, \craig{due to} the use of teacher forcing during the training phase, the model is exposed to the ground-truth word when generating \iadh{the} next word at each time step. However, \iadh{since} there is no ground-truth provided in the testing time, the model \craig{generates} \iadh{the} next word conditioned on its own previous predicted word. \craig{If this is} incorrect, \craig{it may} deviate the whole generated sequence from the actual sequence~\cite{ranzato2015sequence}. This scenario is called exposure bias. To address these issues, we \craig{employ a} reinforcement learning algorithm \craig{that} can directly optimise over the discrete evaluation metric and not rely on the ground truth during training.

\subsubsection{Reinforcement learning training process}

We formulate our query reformulation task as a reinforcement learning problem and employ the REINFORCE~\cite{williams1992simple} algorithm in this work.  The sequence to sequence model acts as the {\em agent}, the parameter $\theta$ of the agent is regarded as the policy $\pi_\theta$ and an action $\hat{y}_{t}$ refers to \iadh{the prediction of} the next word at each time step $t\in[0,M]$. A reward $r(\hat{y_1},\hat{y_2},...,\hat{y_M})$ is observed at the end of the sequence but \craig{is set to} zero when selecting a word within the sequence. The goal of the training is to optimise the policy by maximising the expected reward or minimising the negative expected reward:
\begin{equation}
    \mathcal{L}(\theta)_{RL} = -\mathbb{E}_{\hat{Y} \sim \pi_{\theta}(\hat{Y})}[r(\hat{Y})]
    \label{equ:RLloss}
\end{equation}
where $\hat{Y}=[\hat{y}_{1}, \cdots, \hat{y}_{M}]$ is the predicted sequence and $r(\hat{Y})$ is the observed reward given by the reward function. \xiao{The gradient of Equation~\eqref{equ:RLloss} is \iadh{provided as follows}:}
\begin{equation}
    \nabla_{\theta} \mathcal{L}(\theta)_{RL}=-\mathbb{E}_{\hat{Y} \sim \pi_{\theta}(\hat{Y}) }[r(\hat{Y}) \nabla_{\theta} \log p_{\theta}(\hat{Y})]
\end{equation}

\xiao{In practice, the expectation is estimated using a single sample sequence from the distribution of actions produced from the agent. However, this would cause \iadh{a} high-variance for the model. \iadh{Hence}, a baseline ${r}_{b}$ reward is used to encourage the model to select \iadh{a} sequence with reward $r > r_b$ and discourage those action sequences with reward $r<r_b$. }
\xiao{The gradient of the loss function is as \iadh{follows}:}
\begin{equation}
    \nabla_{\theta} \mathcal{L}(\theta)_{RL} = - \mathbb{E}_{\hat{Y} \sim \pi_{\theta}(\hat{Y}) }[\nabla_{\theta} \log \pi(\hat{Y})(r(\hat{Y})-r_{b}) ]
    \label{equ:RLloss2}
\end{equation}
\xiao{where ${r}_{b}$ is the baseline reward. The baseline $r_b$ \iadh{can be} any estimator that \iadh{is} independent of the action, thus it can reduce the variance of the gradient loss without changing the gradient value (\iadh{since} the second component of Equation~\iadh{(\ref{equ:RLloss2})} can be \craig{proven} to be zero~\cite{rennie2017self}). In this work, we adopt the Self-Critic~\cite{rennie2017self} REINFORCE model, which produces a baseline based on the output at the time of inference rather than estimating the baseline using samples from the current model.}
Another problem for training the RL model is that the action space is very big thus making the model difficult to learn with an initial random policy. \xiao{To avoid starting with an initial random policy, we train the model using the combination of the $\mathcal{L}(\theta)_{ML}$ and $\mathcal{L}(\theta)_{RL}$ loss \xiao{function,}
as follows:}
\begin{equation}
    \mathcal{L}_{train}=\mathcal{N}_{ML} \mathcal{L}(\theta)_{ML}+\mathcal{N}_{RL} \mathcal{L}(\theta)_{RL}
\end{equation}
\xiao{where we first train the model using $\mathcal{L}(\theta)_{ML}$ for $\mathcal{N}_{ML}$ epochs, then train $\mathcal{N}_{RL}$ epochs using  $\mathcal{L}(\theta)_{RL}$~\cite{chen2018keyphrase}.}




\subsection{Reward Function}\label{subsec:reward function}
\xiao{To force our model to learn how to reformulate the input queries into \iadh{a} form that would \iadh{perform} well in the retrieval task, at the end of each predicted sequence, we give a reward \iadh{through} the reward function.}
The reward function for our model is the weighted sum of two components namely, the F1 reward and the QPP reward.

\subsubsection{F1 reward}
To encourage the model to generate an accurate reformulated query, our reward function encapsulates sequence classification accuracy, specifically \iadh{an} F1 reward, therefore \craig{encapsulating \iadh{both} recall and precision \iadh{for the} correct terms}. Recall measures how well the agent could generate identical terms with the ground-truth reformulation, and precision measures how well the agent rejects incorrect words. \craig{In short, the F1 reward encourages the model to} generate \iadh{the} correct 
form of \iadh{a} reformulated query compared to the ground-truth \xiao{paraphrased query}. \craig{However, in our initial experiments, we observed that} seq2seq tends to generate repeated words for our task. Thus, we adopt the 
technique from~\cite{chen2018keyphrase} to penalise the generated sequence by replacing the repetitive words with \craig{the}  $\langle PAD \rangle$ token. Thus the duplicated words are treated as an incorrect generation. 


\subsubsection{QPP reward}
While the F1 reward \craig{aims} to encourage the model \craig{to generate a reformulated query that is close to the ground truth examples (c.f.\ instances in $Y$) in the training dataset, we also want the learned model to generate queries that are likely to be effective in nature.}  To this end, we propose \iadh{the} integration of a query performance \iadh{predictor} into the reward function, \craig{as a signal to} encourage the model to \craig{reformulate} the query from the perspective of improving the \iadh{retrieval} effectiveness. Depending on the \iadh{deployed predictor}, this may guide the reward function to avoid words that are too \iadh{non-informative}.

\craig{It would be possible to integrate a retrieval component into the reward function, and therefore calculate {\em post-retrieval} query performance predictors, which are known to be more accurate~\cite{carmel2010estimating}. However, repeated invocation of the search engine would dramatically slow down the training process. For this reason, we focus on pre-retrieval predictors. We discuss the used predictors later in Section~\ref{ssec:qpp}.}
\craig{Our final reward function is therefore a linear combination of F1 (representing \iadh{the} paraphrase accuracy) and query performance prediction (representing the likelihood that the generated query will be useful to the search engine):}
\begin{equation}
    r(\hat{Y})=\lambda r^{F_1}+(1-\lambda) r^{QPP} \label{reward_function}
\end{equation}
where $\lambda \in[0,1]$ is a \craig{tunable} hyper-parameter \craig{that adjusts the importance of the QPP values within the reward function}. \craig{We assume a default value of $\lambda=0.5$, but investigate the impact of this setting later in Section~\ref{sec:results}.}


\section{Research Questions}\label{sec:RQ}
\looseness -1 \craig{In this work, we \iadh{address} four research questions. Firstly, one of our key contributions 
is the introduction of pre-retrieval query performance predictors (QPPs) for use within the reinforcement learning reward function. In doing so, we assume that they can differentiate between high and low quality query reformulations, to guide the learning process. However, no work has yet investigated QPPs on the MSMARCO dataset, where the queries are question-like in nature. For this reason, we pose our first research question as:}

\textbf{RQ1:} \craig{How accurate are pre-retrieval query performance predictors on the MSMARCO dataset at (a) discriminating between easy and hard queries, and (b) discriminating between high and low quality query reformulations?}


\craig{Secondly, we investigate the effectiveness of \xiao{our proposed DRQR model,}
as follows:}

\textbf{RQ2:} \iadh{Do} queries reformulated using our RL model result in effectiveness improvements over \craig{text generation baselines for generating query reformulations}? 

\looseness-1 \craig{Thirdly, we examine how the effectiveness of the used retrieval approach impacts the effectiveness of our RL model, as \iadh{follows}:}

\textbf{RQ3:} \craig{Does our DRQR model result in further improvements when combined with other} enhanced retrieval approaches such as QE or BERT?


\section{Experimental Setup}\label{sec:exp}
\xiao{In the following, we \iadh{describe} the \iadh{used} MSAMRCO dataset in Section~\ref{ssec:dataset}. \iadh{We discuss our} experimental setup \xiao{for seq2seq and retrieval pipeline} \iadh{in detail} in Section~\ref{ssec:exp_Set} \xiao{and Section~\ref{ssec:exp_ret}}. \iadh{The descriptions of the \xiao{seven} deployed} query performance predictors and \xiaow{\iadh{that of the} four baseline reformulation models are provided in Section~\ref{ssec:qpp} and Section~\ref{ssec:expsetup:baselines}, respectively. Finally, the measures used in our experiments are detailed in Section~\ref{ssec:measures}.}}



\subsection{Dataset}\label{ssec:dataset}
\looseness -1 \craig{All of our experiments are conducted using the MSMARCO document ranking dataset\footnote{\url{https://microsoft.github.io/msmarco/}}, in the setting of the TREC 2019 Deep Learning \iadh{(DL)} track~\cite{craswell2020overview}. In particular, in the TREC \iadh{DL} setting, the corpus is composed of $\sim$3.2M documents, \iadh{along} which are provided $\sim$367k training queries with one or two known relevant documents.}

\craig{In order to train the model to learn how to reformulate \iadh{queries}, we use the training corpus for identifying pairs of queries. In particular, following Zerveas et al.~\cite{zerveasbrown}, we find that some documents \iadh{are} labeled as \iadh{relevant} for multiple queries. We assume that the information needs for such pairs \iadh{of} queries sharing a relevant document are close enough that they can be considered as paraphrases. We identified \iadh{188,292} pairs of such rephrased queries. We sample $90\%$ of the generated pairs as training data, while the remainder $10\%$ is taken as \xiaow{a} validation dataset.}

Finally, to test retrieval effectiveness, we use the 43 new test queries from \iadh{the} TREC Deep Learning Track 2019, which \iadh{were the object of} deep pooling and relevance \iadh{assessments} \xiao{with an average of 153.4 relevant documents per query\iadh{.}}





\subsection{Seq2Seq Setup}\label{ssec:exp_Set}
\looseness -1 For the implementation of the sequence to sequence query reformulation model, we \craig{follow} the setting of Chen et al.~\cite{chen2018keyphrase}, \iadh{where} the hidden size of \iadh{the} encoder and decoder is set as 300. The \xiao{parameters of the model}
are initialised using a uniform distribution \xiaow{--} \craig{i.e.\ we do not use any trained embedding representation.} In the training process, the dropout rate is 0.1 and a gradient clipping of 1.0 is used. In the maximised-likelihood training process, teacher-forcing is applied and the Adam optimiser with a learning rate of $1 \times 10^{-3}$ and a batch size of 12 is used. We also employ \iadh{the early} stopping \iadh{mechanism}, if there are no validation improvements for three \craig{consecutive} \xiao{epochs.}

After obtaining the pre-trained ML model, we \iadh{use it for training} our \xiao{DRQR} 
model. \iadh{The} Adam optimiser with a batch size of 32 and a learning rate of $5 \times 10^{-5}$ is used to train the model. A similar 
early-stopping mechanism \xiao{used in seq2seq setup} is used to early terminate the training. In the decoding phase, we use the greedy search algorithm to generate the reformulated query. Before \iadh{obtaining} the evaluation \iadh{scores} of F1, we remove all the duplicated terms from the prediction sequence~\cite{chen2018keyphrase}. For calculating the pre-retrieval query performance predictor \iadh{scores}, we apply the Porter Stemmer to each token \iadh{since} the index we are using is \iadh{a} stemmed MSMARCO index. For the implementation of the Transformer model, we employ the OpenNMT~\cite{opennmt} platform.


\subsection{Retrieval Pipeline Setup}\label{ssec:exp_ret}

We index MSMARCO using the Terrier IR platform~\cite{ounis2006terrier}, removing standard stopwords and applying Porter stemming. For \iadh{the} retrieval experiments, we make use of the recent Python bindings for Terrier, namely PyTerrier\footnote{\url{https://github.com/terrier-org/pyterrier}}. Our ranking pipeline incorporates \xiao{DPH} 
as well as a BERT re-ranker from the CEDR toolkit~\cite{dai2019deeper}. \iadh{Following} the experimental setup of Su et al.~\cite{suuniversity}, we train the BERT model using 1000 queries from the MSMARCO training dataset ranked to depth 1000. We use 200 queries ranked to depth 50 for \iadh{the} validation of the BERT model; we \iadh{adopt an} early \iadh{termination of the} training process if no further effectiveness improvements are observed on the validation set for 20 epochs. 

\craig{Finally, for all reformulation approaches, we combine the reformulated queries with the original query before retrieval. In doing so, we use a mixing parameter, $\theta$ that controls the influence of the reformulated query, as follows:}
\begin{equation}
q' = q0 + \theta q_r
\end{equation}
\craig{where $q'$ is the final query, \xiaow{$q$0} is the initial query, and $q_r$ is a reformulation. We set the value of $\theta$, as well as the reward tradeoff parameter $\lambda$ in DRQR, by grid searching to maximise the \xiao{NDCG@10}
using a validation set of 200 queries selected from the MSMARCO training set.} 
\subsection{Query Performance Predictors}\label{ssec:qpp}
Our experiments compare \xiao{seven} pre-retrieval query performance predictors~\cite{he2006query,carmel2010estimating,macdonald2012usefulness} from \xiao{five} families:

\paragraph{Inverse Document Frequency (IDF)} 
    
    The inverse document frequency is \iadh{a widely used heuristics} \iadh{for measuring} the relative importance of \iadh{the} query terms \iadh{in a} document. Higher \iadh{IDF} values indicate \iadh{that a} term is infrequent and \iadh{helps to guide the retrieval process}.
    \begin{equation}
        idf(t)=\log \left(\frac{N}{N_{t}}\right)
    \end{equation}
    where $N$ is the number of documents in the whole collection and $N_t$ is the number of documents containing the query term $t$.
    
    \paragraph{Inverse Collection Term Frequency (ICTF)} 
    
    \xiao{Similar to IDF, the inverse collection term frequency measures the relative importance of \iadh{a} query term in \iadh{the} collection D,} \craig{as follows:}
    \begin{equation}
        ictf(t)=\log \left(\frac{|D|}{tf(t,D)}\right)
    \end{equation}
    \looseness -1 \xiao{where $|D|$ is the number of terms in the collection $D$ and $tf(t,D)$ is the \craig{number of occurrences of term $t$ in the whole collection $D$.}} 

    \paragraph{Simplified Clarity Score (SCS)} 
    
    The simplified clarity score measures the Kullback-Leibler Divergence (KL divergence) between the distribution of the \iadh{term in the} query and \iadh{in} the collection D. 
    \begin{equation}
    S C S(q)=\sum_{t \in q} \operatorname{Pr}(t | q) \log \left(\frac{\operatorname{Pr}(t | q)}{\operatorname{Pr}(t | D)}\right)
    \end{equation}
    where $\operatorname{Pr}(t | q)=\frac{tf(t,q}{|q|}$ is the probability of a query term in the query, and $\operatorname{Pr}(t | q)=\frac{tf(t,D)}{|D|}$ is the probability of a query term in the whole collection \iadh{D}.

    \paragraph{Collection Query Similarity (SCQ)} 
    
    The collection query similarity measures the query similarity to the collection. \iadh{\xiaow{A} higher} similarity potentially indicates more \iadh{relevant} documents. 
    \begin{equation}
        SCQ(t)=(1+\log (tf(t, D))) \cdot idf(t)
    \end{equation}
    In \iadh{particular}, the MaxSCQ, AvgSCQ and SumSCQ \iadh{scores} are calculated \craig{by respectively taking the maximum, average or summation of the SCQ scores over the query terms.}
    
    
    \paragraph{Query Length} 
    The number of tokens of a given query. The premise is that longer queries are better specified, and hence \iadh{are likely to} have  \iadh{a} higher effectiveness.
    
\xiao{\iadh{Since} $idf(t)$ and $ictf(t)$ as well as $SCQ(t)$ \craig{are term-level statistics, to obtain query-level effectiveness predictions, we take the average of the statistics over the query terms, and denote these as AvgIDF, AvgICTF and AvgSCQ}. Moreover, following \cite{carmel2010estimating}, we also calculate MaxSCQ and SumSCQ.}


\subsection{Baseline Reformulation Models}\label{ssec:expsetup:baselines}
\xiao{In order to test the \craig{effectiveness} of our \xiao{DRQR} model in generating reformulated queries, we compare our model with various \craig{reformulation} baselines, namely:}


\paragraph{Transformer} \xiao{The transformer model is proposed by \craig{Vaswani} et al.~\cite{vaswani2017attention}. \craig{Following the} setup of Zerveas et al.~\cite{zerveasbrown}, we \craig{use the} OpenNMT platform~\cite{opennmt}. \craig{In~\cite{zerveasbrown}}, the authors \iadh{generated} three rephrased queries and concatenated these to the original query to form a new query. We apply the Transformer model with one, three and five generated paraphrases obtained using \iadh{the} Beam Search technique in the decoding phase. \iadh{These} are denoted as Transformer$_{1}$, Transformer$_{3}$ and Transformer$_{5}$, respectively. } 

\paragraph{Sequence to Sequence Model with Attention} This sequence-to-sequence model of~\cite{bahdanau2014neural}, including attention \iadh{is} used by~\cite{sutskever2014sequence}. \craig{This baseline is again implemented using the} OpenNMT platform~\cite{opennmt}. 
 
\paragraph{CatseqML Model} Compared to the previous model, CatseqML adds the copy mechanism (Equation~\eqref{eqn:copy}). CatseqML is trained using \iadh{the} maximum-likelihood loss function~\cite{yuan2018one}. In this baseline, the original query is regarded as the input source text, the ground-truth paraphrase is taken as a set of one-word keyphrases.

\looseness -1 \paragraph{CatseqRL Model}  \xiao{Compared to CatseqML, CatseqRL is trained using reinforcement learning~\cite{chen2018keyphrase}. The reward only \xiao{uses} the F1 score calculated from the predicted sequence and input sequence. \craig{Hence, this model is identical to Equation~\eqref{reward_function} with $\lambda=1$, i.e.\ without considering any query performance predictors in the reward function.} \craig{As for} CatseqML, the ground-truth paraphrase is regarded as a set of one-word keyphrases extracted from the input text.}

\subsection{Measures}\label{ssec:measures}

\craig{Our experiments encapsulate two types of \iadh{measurements}, \iadh{as follows}: for measuring retrieval {\em effectiveness}; and \iadh{for} measuring  the {\em accuracy} of the query performance predictors. In particular, for measuring effectiveness we make use of mean average precision (MAP) and normalised discounted cumulative gain (NDCG@10), which were the official measures reported in the TREC 2019 Deep Learning track overview~\cite{craswell2020overview}. We use the paired t-test for testing significant differences between effectiveness values.}

\craig{For measuring the QPP accuracy, we rank queries based on the QPP \iadh{values}, as well as by retrieval effectiveness, and then compute rank correlation coefficients. In particular, following~\cite{carmel2010estimating}, we compute Spearman's $\rho$ correlation and Kendall's $\tau$ rank correlation \xiaow{--} a high absolute correlation for a given predictor indicates that the predictor \iadh{accurately} predicts \iadh{the} performance. To determine if a correlation is significant, we \xiao{perform \iadh{a} \xiao{permutation test}};
 we determine if the differences \iadh{between} two correlations are significantly \iadh{different} using a Fisher-z transform.}

\section{Results}\label{sec:results}
\looseness -1 \xiao{In the following, we present our \craig{findings for RQ1 concerning \iadh{the} QPP accuracy} in Section~\ref{ssec:RQ1}.} \craig{Findings for the effectiveness of DRQR viz.\ RQ2 and RQ3 are reported in Sections~\ref{ssec:RQ2} and~\ref{ssec:RQ3} respectively.}

\subsection{RQ1: Query Performance Predictors}\label{ssec:RQ1}

\begin{table*}[tb]
\caption { Correlation between different QPP predictors \craig{and \iadh{the} retrieval evaluation measures}. \xiao{The strongest correlation is emphasised. The $*$ symbol  denotes a significant correlation between \craig{the predictor and the retrieval measure} ($p<0.05$), \craig{while the} $<$ symbol denotes a significant degradation from the best predictor in that column ($p<0.05$), \craig{according to a Fisher-z transform}.} \xiao{The left-hand side of the table presents the correlation analysis on the 43 TREC queries, while the right-hand side is the correlation analysis conducted on the 4*43 reformulated queries \craig{obtained from four query reformulation} baselines.}}

\begin{tabular}{l|cccc|cccc}
 \hline \hline 
 & \multicolumn{4}{c|}{Queries ($N=43$)} & \multicolumn{4}{c}{Query Reformulations ($N=43*4 = 172$)} \\
 \multirow{2}{*}{QPP predictors}  & \multicolumn{2}{c}{Spearman's $\rho$}  & \multicolumn{2}{c|}{Kendall's $\tau$} & \multicolumn{2}{c}{Spearman's $\rho$}  & \multicolumn{2}{c}{Kendall's $\tau$}  \\ 
  \cline{2-9} 
	&  MAP  & NDCG@10                                                 
	&  MAP  & NDCG@10  
	&  MAP  & NDCG@10 
	&  MAP  & NDCG@10    \\    \hline

  AvgIDF & 0.431*   & 0.443*        
    & 0.318*   & 0.348*  
    & 0.305*   &0.325*
    & 0.231*	& 0.236*    \\  
 
 SCS   & 0.442*   & 0.434*       
   & 0.318*   & 0.318*   
   & 0.230*	& 0.283*
   & 0.169*	& 0.191*   \\ 
 
 AvgSCQ   & \textbf{0.464*}   & 0.440*    
    & \textbf{0.324*}   & 0.311*  
    & \textbf{0.371}*	& \textbf{0.383}*
    & \textbf{0.267}*	& \textbf{0.270}*     \\ 
 
 AvgICTF   & 0.443*   & \textbf{0.469*}        
    & \textbf{0.324*}   & \textbf{0.360*}    
    & 0.249*	& 0.269*
    & 0.187*	& 0.181*  \\ 
 
 MaxSCQ   & 0.211      & 0.234       
    & 0.139     & 0.185     
    & 0.129	& 0.204
    & 0.104	& 0.159 \\ 
 
 SumSCQ   & 0.157     & 0.129      
    & 0.110     & 0.0742   
    & 0.202	& 0.087$<$
    & 0.130  & 0.0536$<$  \\ 
 
 QueryLength   & $-0.0162$$<$   & $0.0114$$<$        
    & $-0.0131$$<$   & -0.00358  
    & 0.0343$<$   &	$-0.0908$$<$
    & 0.0152$<$   &	$-0.0764$$<$  \\ 
 \hline \hline
\end{tabular}
\label{tab:corr}
\end{table*}

\begin{figure}
    \centering\vspace{-\baselineskip}
    \includegraphics[width=3in]{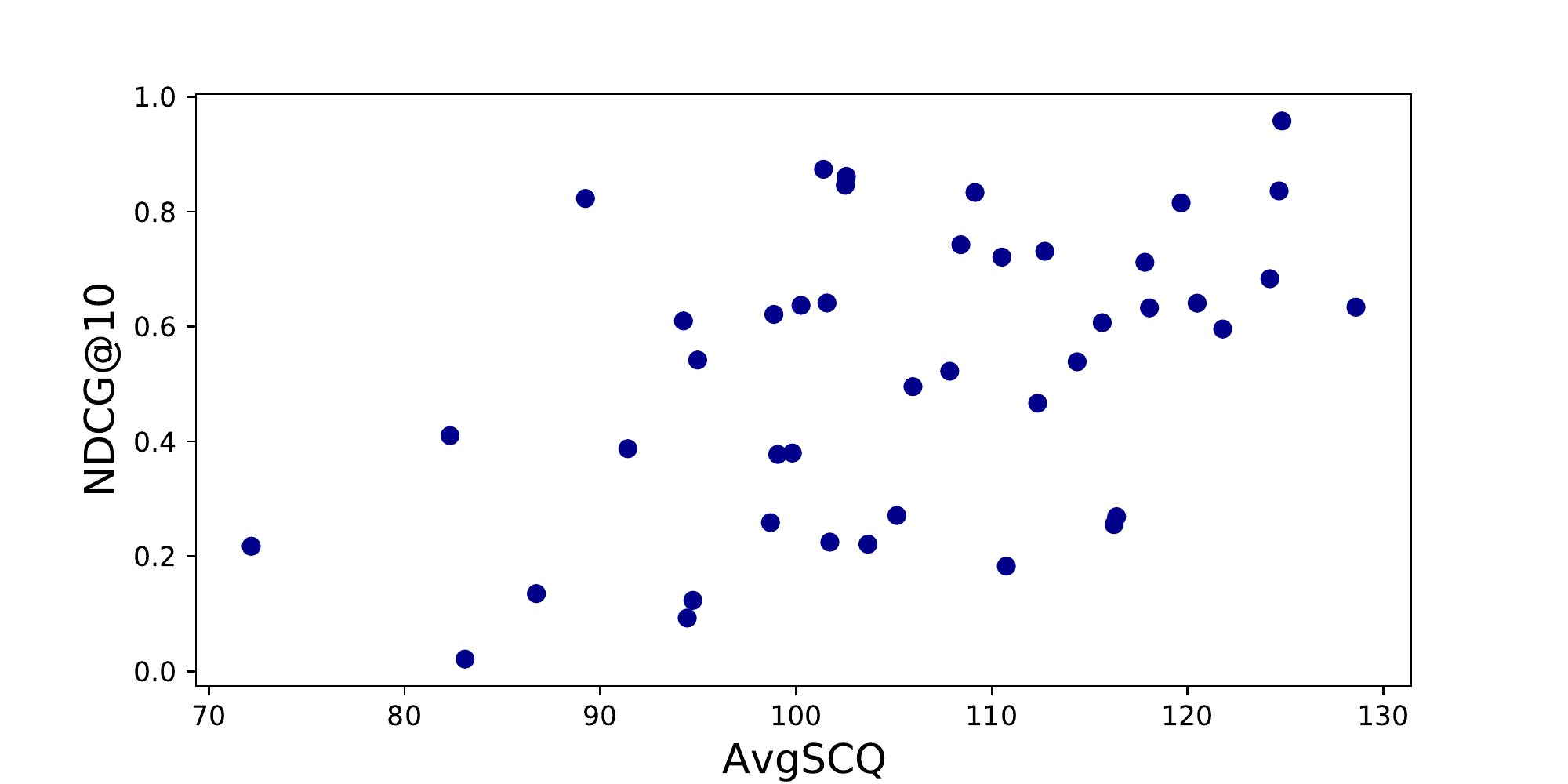}
    \caption{An example of the correlation between \iadh{a} QPP predictor AvgSCQ value (x-axis) and NDCG@10 (y-axis). Each point denotes a TREC test query. }
\label{fig:scatter}\vspace{-\baselineskip}
\end{figure}

\craig{In this section, we investigate the accuracy of \iadh{the} pre-retrieval query performance predictors on the MSMARCO document ranking dataset, both for predicting the effectiveness of queries, as well as \iadh{for} different reformulations.}


\looseness -1 \craig{Firstly, to illustrate \xiao{the correlation between a particular QPP predictor and \iadh{an} IR effectiveness metric, e.g. NDCG@10,} Figure~\ref{fig:scatter} contains \xiaow{a scatter plot showing how the predictions of the AvgSCQ QPP \iadh{scores} are correlated with the NDCG@10 retrieval effectiveness.}
Each point denotes a particular query among the $N=43$ TREC queries.} \xiao{The x-axis of each point is the calculated QPP predictor value for the query, while the y-axis of the point is the value of its NDCG@10 \iadh{performance}. \craig{The more that \iadh{the} points fall on the principal diagonal}, \iadh{the stronger} the correlation between the predictor and \iadh{the} NDCG@10 performance. In contrast, irregular \iadh{and other} dispersed points \iadh{denote a} weak correlation.} \craig{To quantify \xiaow{the observed correlation}, the left hand side of Table~\ref{tab:corr} contains the Spearman's $\rho$ and Kendall's $\tau$ \iadh{correlations} between different QPP predictors and ranking effectiveness \iadh{metrics} (namely mean average precision and NDCG@10). All correlations are calculated on the $N=43$ TREC queries. In the table, \iadh{the} highest correlations in each column are emphasised, and significant correlations are denoted with $*$. We observe that AvgSCQ exhibits the highest $\rho$ for MAP (0.464), while under $\tau$, AvgICTF and AvgSCQ \craig{have identical correlations for MAP (0.324)}. Overall, we observe medium (and significant) correlations (0.3-0.4) for most of the QPPs except MaxSCQ and SumSCQ, and indeed, AvgIDF, SCS, AvgSCQ, and AvgICTF are statistically indistinguishable among themselves.}

We now consider these results in the context of historical performances of pre-retrieval predictors reported in the literature. He \& Ounis~\cite{he2006query} found SCS and AvgIDF to be the most accurate \xiao{predictors} on the TREC Robust track queries, observing correlations with $\rho \approx 0.4$; Carmel~\cite{carmel2010estimating} reported similar observations concerning the accuracy of SCS and AvgIDF on Robust, WT10G and GOV2 test collections; later, Tonellotto \& Macdonald~\cite{tonellotto2020using} only observed $\tau$ correlations $<$ 0.25 on 200 TREC Web track queries calculated on the TREC ClueWeb09 test collection, but observed SCQ to be among the most accurate pre-retrieval predictors. Our results demonstrate that \xiao{pre-retrieval} query performance predictors are more accurate on MSMARCO than on the shorter Web track queries, and mirror previous observations on older test collection such as Robust. Hence, in answer to RQ1(a), we find that \iadh{four of the used} pre-retrieval predictors exhibit medium but significant correlations on the 43 TREC queries \iadh{using the MSMARCO dataset}.


\craig{Our use of QPPs within the reinforcement learning reward function assumes that they can differentiate between good and bad query reformulations. To test this, instead of assessing the accuracy of the predictors for the original queries, we now assess their effectiveness at ranking query reformulations. In particular, for each of the 43 TREC queries, we consider the reformulations made by the \xiao{four  baseline reformulation models, namely Seq2seq with attention, Transformer$_{1}$ (i.e. the Transformer model with one generated sequence), CatseqML and \iadh{the} CatseqRL model). This gives a total population of $N=43 * 4 = 172$ query reformulation instances; }we obtain the \iadh{predictors'} values for each reformulation instance, and measure the correlation with the effectiveness of the reformulation.} \xiao{The results are presented in the right hand side of Table~\ref{tab:corr}}.

\looseness -1 \craig{On analysis of the right hand side of \iadh{Table~\ref{tab:corr}}, we observe that, in general, the QPPs are able to differentiate between good and bad reformulations, \iadh{since} significant 
correlations \xiao{under the permutation test} are observed, which are only slightly lower than those observed in the left hand side of the table.} Moreover,  the four best predictors from the left hand side \iadh{of the table}, \iadh{namely} AvgIDF, SCS, AvgSCQ and AvgICTF, are still the best \iadh{predictors} \iadh{using} the reformulations, and are statistically indistinguishable among each other. The low-performing predictors from the left-hand side of \iadh{Table~\ref{tab:corr}}, \iadh{namely} SumSCQ, MaxSCQ, QueryLength, remain inaccurate. This answers RQ1(b) \xiao{that the pre-retrieval predictors can distinguish between high and low quality query reformulations.} 
For this reason, we take forward these four predictors into our experiments for research question RQ2.

\subsection{RQ2: DRQR vs.\ reformulation baseline models?}\label{ssec:RQ2}

\begin{table}[tb]
    \centering
    \caption{Comparison between the DRQR model and the query reformulation baselines. \xiao{The symbol * denotes a significant difference between the current query reformulation model and the query reformulation model \iadh{that} achieves the best performance with the same ranking model and the same effectiveness metric (paired t-test, $p<0.05$)}.}  \label{tab:com_baselines}
    \resizebox{85mm}{!}{
    \begin{tabular}{cccc}
    \hline \hline Query Reformulation Model & Ranking Model & MAP & NDCG@10 \\
    \hline \hline \multirow{2}{*} { Transformer$_{1}$ } & DPH & 0.2378* & 0.3712*\\
& BM25 & 0.2467* & 0.3438* \\
    \hline \multirow{2}{*} { Transformer$_{3}$ } & DPH & 0.1606* & 0.2613*\\
& BM25 & 0.1648* & 0.2471* \\
    \hline \multirow{2}{*} { Transformer$_{5}$ } & DPH & 0.1287* & 0.2065*\\
& BM25 & 0.1363* & 0.1983* \\
    \hline \multirow{2}{*} { Seq2seq$_{attention}$ } & DPH & 0.2907* & 0.4557* \\
& BM25 & 0.2907* & 0.4350* \\
    \hline \multirow{2}{*} { CatseqML } & DPH & 0.2999*& 0.4795* \\
& BM25 & 0.3160 & 0.4754* \\
    \hline \multirow{2}{*} { CatseqRL } & DPH & 0.3125 & 0.5156 \\
& BM25 & \textbf{0.3465} & 0.5018 \\
     \hline \multirow{2}{*} { DRQR (AvgSCQ) } & DPH & \textbf{0.3293} & \textbf{0.5516} \\
& BM25 & 0.3316 & \textbf{0.5467} \\
    \hline \hline
    \end{tabular}}
  
\end{table}

\craig{Next, we examine the effectiveness of the text generation query reformulation models, including our proposed DRQR model, and those listed in Section~\ref{ssec:expsetup:baselines}.}  Table~\ref{tab:com_baselines} presents the effectiveness of the different reformulation models, when applied to either the DPH or BM25 retrieval models. In this table, DRQR uses the AvgSCQ predictor, along with the \craig{default} reward tradeoff parameter $\lambda=0.5$ in Equation~\eqref{reward_function} \xiaow{--} later, we revisit each of these choices. \craig{Further, for each reformulation model, we append the generated query reformulations with the corresponding original query \xiaow{--} as the reformulated query alone is not \iadh{sufficiently} effective}~\cite{zerveasbrown}. 
Within Table~\ref{tab:com_baselines}, the best result \craig{in each column is} highlighted in bold and the symbol $\ast$ denotes a \iadh{significant} \craig{degradation} \iadh{of} the best result, according to the paired t-test for $p<0.05$.


\craig{On analysis of  Table~\ref{tab:com_baselines}, we observe that the baseline reformulation models, namely the Transformers models, as well as seq2seq with attention, \iadh{and} CatseqML or CatseqRL, do not generate effective reformulations. Indeed, recall that Transformer$_{3}$ corresponds to the existing approach of Zerveras~\cite{zerveasbrown}. In contrast, our proposed DRQR model \iadh{outperforms} these models in terms of both MAP and NDCG@10. These improvements are significant (paired t-test, $p<0.05$) over all reformulation models except CatseqRL (one exception \iadh{being} CatseqML for BM25 on MAP). The effectiveness of CatseqRL over the other models supports the benefit of reinforcement learning to avoid the exposure bias problem (discussed earlier in Section~\ref{ssec:related:RL}).}

\looseness -1 Furthermore, our approach exhibits marked but not significant improvements over CatseqRL -- for instance, DRQR exhibits a 6.9\% improvement in NDCG@10 for DPH (0.5156 $\rightarrow$ 0.5516).  We argue that this is because our model has the ability to avoid generating queries that are predicted not to be effective, while  traditional text generation models are focused instead on generating correct paraphrases, \iadh{where} they may \iadh{consequently} exhibit \iadh{a} topical drift away from the user's original information need, thereby damaging effectiveness. 

\begin{figure}[tb]
    \centering
    \includegraphics[width=\linewidth]{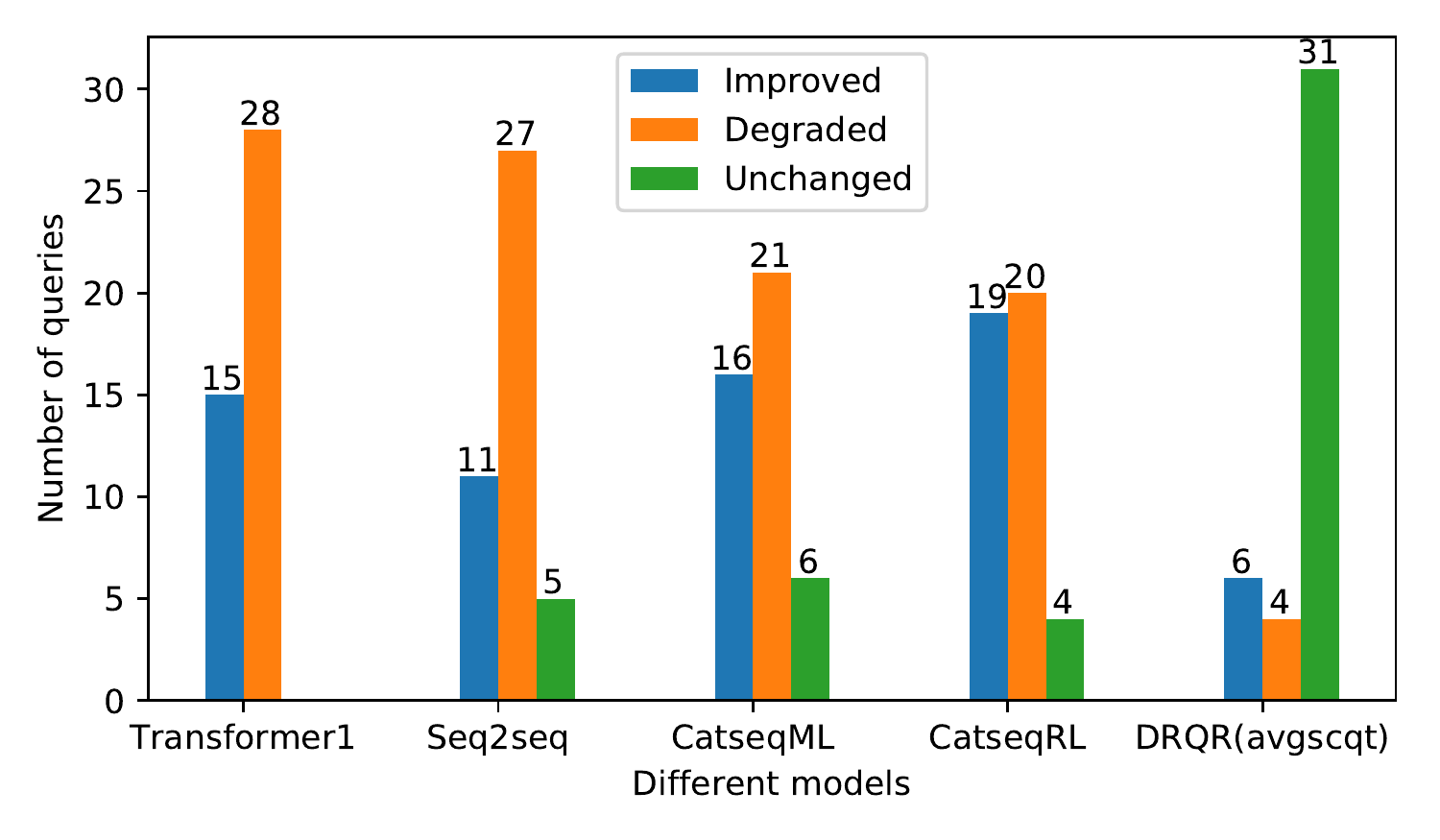}\vspace{-\baselineskip}
    \caption{Histogram of improved/degraded/unchanged number of queries for each query reformulation model.}
    \label{fig:histogram}
\end{figure}

\looseness -1 We further \craig{examine the} \iadh{performances} on a per-query basis for the Transformer$_{1}$, Seq2Seq$_{attention}$, CatseqML, CatseqRL and DRQR models. Figure~\ref{fig:histogram} compares the number of improved, degraded and unchanged queries for the query with and without reformulated \iadh{queries in terms of} NDCG@10 given by the DPH ranking model. In Figure~\ref{fig:histogram}, we can see that while our proposed DRQR \iadh{model does not} possess the largest number of improved queries, it has the least number of degraded \iadh{queries}, and many unchanged queries. The reason behind this is that the query performance prediction in our DRQR model has an effect of penalising words that might downgrade the retrieval performance. In addition, \xiao{Table~\ref{tab:perQuery} shows three reformulated queries with improved \iadh{performances} over \iadh{their corresponding} raw query for each query reformulated model.}
\iadh{We} can see that \iadh{the} paraphrase models tend to reformulate an input query into a question-type query beginning \iadh{with} ``what is'' or ``how''. 


\craig{Finally, we return to address the choice of query performance predictor within DRQR. Table~\ref{tab:com_dif_qpp} reports the effectiveness of the DRQR models applying the four best QPPs from Section~\ref{ssec:RQ1}. From the table, we observe that while AvgSCQ is the best predictor, there is no significant differences between the effectiveness of the different models, according to a paired t-test. It is also of note that AvgSCQ was the best predictor of reformulation quality in Section~\ref{ssec:RQ1} (see Table~\ref{tab:corr}, right hand side). AvgSCQ considers the similarity between \iadh{the} query terms and the corpus, and hence focuses the DRQR model on generating query terms that are ``frequent but not too frequent" in the collection, thereby both preventing too many \iadh{non-informative} terms being added to the query (as AvgICTF and AvgIDF does), but also ensuring that the terms being added to the query have sufficient documents in the collection.}

\craig{Overall, in response to RQ2, we find that our proposed \iadh{DRQR} model outperforms, significantly, existing text generational models that do not apply reinforcement learning. Moreover, reinforcement learning provides a marked boost in effectiveness, while the introduction of a pre-retrieval query performance predictor to guide the model towards creating queries that appear to be effective, results in further effectiveness improvements.}


\begin{table}[tb]
    \centering
    \caption{\craig{Effectiveness comparison between DRQR models using different QPPs (no significant differences observed according to \iadh{a} paired t-test at $p<0.05$).}}\label{tab:com_dif_qpp} \resizebox{85mm}{!}{
    \begin{tabular}{cccc}
\hline \hline Query Reformulation Model & Ranking model & MAP & NDCG@10 \\
\hline \hline \multirow{2}{*} { DRQR (AvgICTF) } & DPH & 0.2742 & 0.4834 \\
& BM25 & 0.2846 & 0.4578 \\
\hline  \multirow{2}{*} { DRQR (SCS) } & DPH & 0.2804 & 0.4960 \\
& BM25 & 0.2844 & 0.4456 \\
\hline \multirow{2}{*} { DRQR (AvgIDF) } & DPH & 0.2985 & 0.4795 \\
& BM25 & 0.3160 & 0.4754 \\
\hline \multirow{2}{*} { DRQR (AvgSCQ) } & DPH & \textbf{0.3293} & \textbf{0.5516} \\
& BM25 & \textbf{0.3316} & \textbf{0.5467} \\
\hline \hline
\end{tabular}}
\end{table}





\begin{table}[tb]
    \centering
    \caption{Examples of the reformulated queries obtained using different query reformulation models.}\label{tab:perQuery}
     \resizebox{85mm}{!}{
    \begin{tabular}{lllr}
\hline \hline & Original Query & Reformulated Query  \\
\hline \hline \multirow{3}{*} { Transformer$_{1}$ } & what types of food can you cook sous vide & 
 vide what are the of you to for a cook  \\
& cost of interior concrete flooring & how much is the of it to for a concrete floor  \\
&why did the us volunterilay enter ww1 &  what did the of you have in germany \\

\hline  \multirow{3}{*} { seq2seq$_{attention}$} & rsa definition key & 
 what is rsa \\
& definition declaratory judgment & what was teh declaratory act \\
& causes of left ventricular hypertrophy &  what is ventricles  \\

\hline \multirow{3}{*} { CatseqML } & what is physical description of spruce &   what is the of you for a spruce \\
& when was the salvation army founded &  what is the of you for salvation  \\
&define visceral & what is the of your visceral \\

\hline \multirow{3}{*} { CatseqRL } &what is the most popular food in switzerland &  what is the of you for switzerland  \\
& what is durable medical equipment consist of &  what is the of you for dme   \\
&rsa definition key &  what is the of rsa  \\

\hline \multirow{3}{*} { DRQR (AvgSCQ) } & rsa definition key&  what is the rsa of before \\
&types of dysarthria from cerebral palsy & what is the palsy of before something \\
& how to find the midsegment of a trapezoid & what is the trapezoid of before a \\

\hline \hline
\end{tabular}}
\label{tab:examples}
\end{table}

\subsection{RQ3: Does our DRQR approach combine with other enhanced retrieval approaches such as QE or BERT?}\label{ssec:RQ3}

\looseness -1 \craig{In this section, we compare DRQR with other retrieval models, and also experiment to determine if it can be combined with these models. We focus on \iadh{the parameter-free} DPH \iadh{model}, \iadh{since} the observed trends were similar between DPH and BM25 in Section~\ref{ssec:RQ2}. In particular, we use DPH, DPH + Bo1 query expansion~\cite{amati2002probabilistic}, as well as a BERT re-ranker (as implemented by the CEDR toolkit~\cite{macavaney2019cedr}). Retrieval using the original query is denoted as \xiaow{$q$0}. In this section, both the reformulation weight $\theta$, as well as the reward tradeoff hyperparameter $\lambda$ are trained using the validation set. We again apply AvgSCQ as the QPP in DRQR.}
\craig{Table~\ref{tab:adv-RM} reports the effectiveness results, comparing DRQR vs.\ the original query formulation (denoted q0) \iadh{using different ranking models}. From the results, given these experimental settings, we note that DRQR improves NDCG@10 in 3 out of 3 cases, and improves MAP in 2 out of 3 cases. The disparity between MAP and NDCG@10 mirrors some of our earlier findings in \cite{suuniversity}, where we found that MAP and NDCG@10 responded differently on the MSMARCO dataset.} In general, while DRQR is not as effective as query expansion, it can help to enhance the effectiveness of QE.


\looseness -1 \craig{On the other hand, none of the improvements are significant according to a paired t-test; this is because, as shown in Figure~\ref{fig:histogram}, the number of queries altered by DRQR is not \iadh{sufficiently} large; its clear
\xiao{from Figure~\ref{fig:histogram}} that \iadh{the} addition of the QPP component makes the RL model more conservative in nature; moreover, \xiao{\craig{from} Table~\ref{tab:examples}}, both \iadh{the} DRQR and CatseqRL models generate similar reformulations.} \xiao{\craig{Indeed, on closer inspection of the} generated reformulations for the 43 test queries by each query reformulation model, we find that 35/43 queries for DRQR and 28/43 for CatseqRL are reformulated \iadh{into queries that} start with ``what is'', while 
the proportion is 28/43 for CatseqML, 23/43 for seq2seq with attention model and 17/43 for Transformer$_1$ model.} \craig{We postulate that this focus on question-like n-grams are due to the absence of any pre-trained term representations for the text generation. We hope to address this in future work.}



\begin{figure}
    \centering
    \includegraphics[width=85mm]{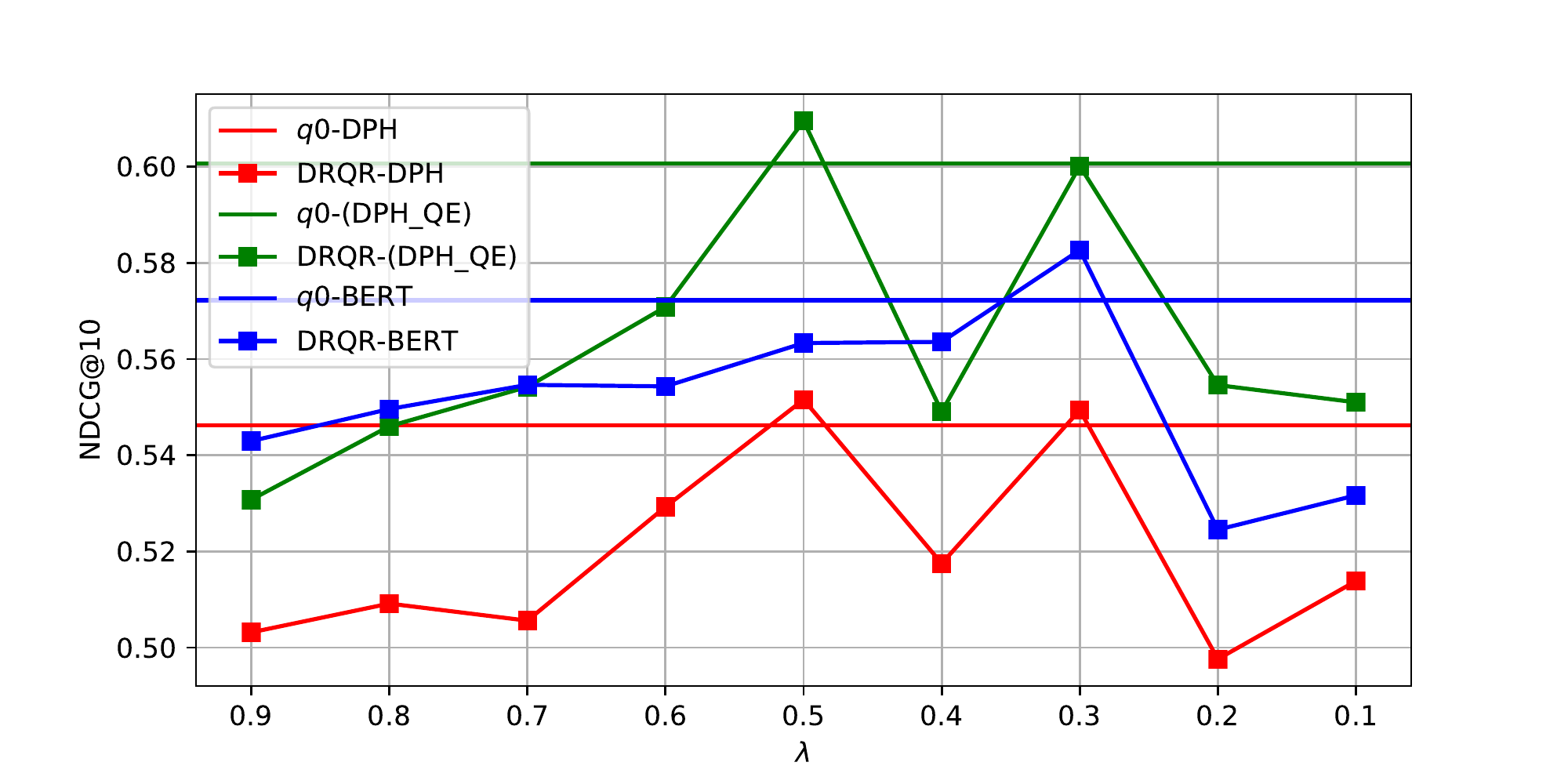}
    \caption{Impact of varying \iadh{the} reward tradeoff parameter $\lambda$. \xiao{Retrieval approaches are grouped by colour.}  }
    \label{fig:rewardtradeoff}
\end{figure}

\craig{We now investigate the impact of the reward tradeoff hyperparameter $\lambda$ from Equation~\eqref{reward_function}. We demonstrate its impact on \iadh{the} NDCG@10 performance in Figure~\ref{fig:rewardtradeoff}}, while holding $\theta=1$. 
From the figure, we observe that $\lambda$ values of 0.5 or 0.3 are the most effective, regardless of the retrieval approach.

Overall, in answer to RQ3, we conclude that our DRQR model demonstrates some \iadh{promising trends}, \iadh{by} improving \iadh{three different} retrieval approaches, \iadh{albeit} not by a significant margin.

\begin{table}[tb]
    \centering
    \caption{\looseness -1 Comparison between different ranking models with and without DRQR (i.e. $q0$ denotes the original query). For each ranking model and measure, the best result is emphasised. }
    \begin{tabular}{llcc}
     \hline \hline & Ranking model & MAP & NDCG@10 \\
     \hline \hline 
     \xiaow{$q$0} & DPH & 0.3332 &  0.5462 \\
     DRQR (AvgSCQ) & DPH & \bf 0.3353 & \bf 0.5470 \\

     \xiaow{$q$0} & DPH+QE & {\bf 0.3992} & { 0.6008} \\
     DRQR (AvgSCQ) & DPH+QE & 0.3989 & \bf 0.6017 \\
     \xiaow{$q$0} & DPH+BERT &  0.2702 & 0.5722 \\
     DRQR (AvgSCQ)   & DPH+BERT & \bf 0.2741 & \bf 0.5773 \\
    \hline \hline
    \end{tabular}
    \label{tab:adv-RM}
    
\end{table}

\section{Conclusions}\label{sec:con}
\craig{In this work, we proposed a deep reinforcement learning text generation model for query reformulation, \iadh{called} DRQR, which includes \iadh{both an} attention and copying mechanisms. \iadh{DRQR also includes the} novel integration an of existing IR technique, \iadh{through} the introduction of pre-retrieval query performance prediction into the reward function. Our experiments on the TREC Deep Learning track test collection demonstrated that pre-retrieval \iadh{query performance predictors} were able to distinguish between both high and low effectiveness queries on this test collection, as well high and low effectiveness query reformulations. Taking these observations forward, we demonstrated that the use of reinforcement learning results in enhanced query reformulations compared to other classical text generation models, and that query performance predictors further result in more effective reformulations. Finally, we integrated DRQR with various retrieval models, and found that it could enhance retrieval effectiveness, but not by a significant margin}. \craig{\iadh{As} future work, we \iadh{aim to} consider the integration of query performance predictors (which are differentiable) as a regularisation directly within non-reinforcement learning models such as CatseqML,} \craig{as well as use of pre-trained embeddings model for text generation, such as T5~\cite{raffel2019exploring}.}

\bibliographystyle{ACM-Reference-Format}
\bibliography{sample-base}

\end{document}